%% file: proceeding.tex
\documentclass[12pt]{article}
\usepackage{graphicx}

\newcommand\pubnumber{NuPhys2017-Guzowski}
\newcommand\pubdate{\today}

\def\manchester{The University of Manchester, School of Physics and Astronomy\\
Oxford Road, Manchester M13 9PL, United Kingdom}
\def\support{\footnote{On behalf of the NEMO-3 Collaboration.}}

\def\Title#1{\begin{center} {\Large #1 } \end{center}}
\def\Author#1{\begin{center}{ \sc #1} \end{center}}
\def\Address#1{\begin{center}{ \it #1} \end{center}}

\newcommand\pubblock{\rightline{\begin{tabular}{l} \pubnumber\\
         \pubdate  \end{tabular}}}
\newenvironment{Abstract}{\begin{quotation}  }{\end{quotation}}
\newenvironment{Presented}{\begin{quotation} \begin{center} 
             PRESENTED AT\end{center}\bigskip 
      \begin{center}\begin{large}}{\end{large}\end{center} \end{quotation}}

\input econfmacros.tex
\begin{document}
\begin{titlepage}
\pubblock

\vfill
\Title{The first ever search for neutrinoless quadruple beta decay}
\vfill
\Author{Pawel Guzowski\support}
\Address{\manchester}
\vfill
\begin{Abstract}
  In some models beyond the Standard Model where lepton number violation by two units is forbidden (implying neutrinos cannot be Majorana fermions), violation by four units can still occur, allowing neutrinoless quadruple beta decay. The isotope $^{150}$Nd is one of the very few isotopes where this process could be experimentally observed. The experimental signature consists of four electrons with a total energy equal to the $Q$-value of the decay. The NEMO-3 experiment in the Modane Underground Laboratory in France ran from 2003 to 2011, and studied seven isotopes of interest including $^{150}$Nd. This article describes the result of the first experimental search for neutrinoless quadruple beta decay using the entire NEMO-3 data set for 36.5~g of $^{150}$Nd with 5.25~years of data. No evidence is found of this decay, and lower limits at the 90\% CL on the half-life are set in the range $(1.1-3.2)\times10^{21}$~years.
\end{Abstract}
\vfill
\begin{Presented}
NuPhys2017, Prospects in Neutrino Physics\\
Barbican Centre, London, UK,  December 20--22, 2017
\end{Presented}
\vfill
\end{titlepage}
\def\thefootnote{\fnsymbol{footnote}}
\setcounter{footnote}{0}

\section{Introduction}

The search for lepton number violation ($\Delta L\neq0$) has a long history and promises to answer fundamental questions about the Dirac or Majorana nature of the neutrino. Any elementary process that violates lepton number by two units ($\Delta L=2$) will imply that the neutrino is its own antiparticle and give rise to a Majorana neutrino mass term in the Lagrangian~\cite{blackbox}. However there has been recent interest in new physics with lepton number violation by more than two units~\cite{theories}, and the Majorana term would be forbidden if the minimum $\Delta L\ge 3$. In particular, new models with $\Delta L=4$ could lead to neutrinoless quadruple beta decay ($0\nu4\beta$)~\cite{qbd}. This article describes the first ever search for this rare decay, which has been performed by the NEMO-3 experiment and published in Ref.~\cite{prl}.

\section{Quadruple beta decay}

In quadruple beta decay, an isotope $(Z,A)$ decays to a daughter isotope with $(Z+4,A)$ releasing four electrons. Under the Standard Model (SM) of particle physics, four electron anti-neutrinos are also released that carry away invisible energy, and the electron energy sum is continuous up to the decay $Q$-value. The expected half-life is extremely long, \hbox{$T_{1/2}\gg 10^{100}$~years}, due to the 8-particle phase space which scales as $Q^{23}$. The neutrinoless mode of this decay, which is beyond the SM, violates lepton number by four units. In this mode, no neutrinos are released in the decay and the kinetic energy of the four electrons will sum to a single value, the $Q$-value of the decay. The four particle phase space scales only as $Q^{11}$ and could give this decay (if it exists) a much shorter half-life than the SM decay. 

For an isotope to be a viable candidate to study this decay, it has to be stable to alpha, beta, and gamma decays, and the daughter isotope has to have less mass for the decay to be energetically allowed. There are only three such known isotopes: $^{136}$Xe, $^{96}$Zr, and $^{150}$Nd, with their properties listed in Table~\ref{tab:isotopes}.
\begin{table}[bp]
\begin{center}
\begin{tabular}{lccc}  
  Isotope & $Q$-value (MeV) & $Q_{2\beta}$ (MeV) & Abundance (\%) \\ \hline
  $^{96}$Zr & 0.63 & 3.34 & 2.8 \\
  $^{136}$Xe & 0.05 & 2.46 & 8.9 \\
  $^{150}$Nd & 2.08 & 3.37 & 5.6 
\end{tabular}
  \caption{The only three viable isotopes for quadruple beta decay, their $Q$-values, the $Q$-value of the isotope's double beta decay ($Q_{2\beta}$), and the isotopic abundance.}
\label{tab:isotopes}
\end{center}
\end{table}
The $Q$-value of $^{150}$Nd is the highest, making it the best isotope to search for this decay, given that the phase-space factor scales as $Q^{11}$, and a signal of a higher $Q$-value will lie above less energetic radioactive backgrounds.

The signal for this process will be four electrons emitted from a decay with kinetic energy that sums to the $Q$-value. Each of the viable isotopes also decays by double beta decay with a $Q$-value ($Q_{2\beta}$ in Table~\ref{tab:isotopes}) larger than that of quadruple beta decay. The $0\nu4\beta$ energy peak lies in the middle of the broad energy spectrum due to SM double beta decays, and so distinguishing these two decays will be impossible for experiments that measure only energy. To unambiguously identify $0\nu4\beta$ decays, reconstruction of all four individual electrons is necessary.

\section{The NEMO-3 experiment}

The NEMO-3 experiment was designed to search for neutrinoless double beta decay and ran from 2003 to 2011 at the \emph{Laboratoire Souterrain de Modane} (Modane Underground Laboratory), at a depth of 4800~m water equivalent under the Frejus mountain, on the French-Italian border. Thin foils of seven different isotopes were analysed. Each of 140 source foils was composed of either a pure metallic isotope, or a composite of isotope powder mixed in PVA glue on a mylar support sheet. The foils were approximately 6~cm wide and 2.5~m tall, and had thickness between 60--300~$\mu$m. Tracking chambers were placed either side of the foils, with a height of $\sim3$~m and a depth of $\sim50$~cm. Calorimeter walls composed of 1940 scintillator blocks coupled to PMTs surrounded the tracker. The tracker had a single-electron vertex resolution of 3~mm in the horizontal direction and 10~mm in the vertical. The calorimeter had an energy resolution of $14\%/\sqrt{E/\textrm{MeV}}$ and 250~ps timing resolution. The tracker, calorimeter, and a 25~G magnetic field allowed for the discrimination between beta electrons, positrons, gamma rays, and alpha particles, and the counting of individual electrons coming from a decay.
Many measurements~\cite{n3} have been made with $^{100}$Mo, $^{82}$Se, $^{130}$Te, $^{116}$Cd, $^{48}$Ca, $^{96}$Zr, and $^{150}$Nd.
 
\section{Searching for the neutrinoless quadruple beta decay of $^{150}$Nd}

The SM double beta decay half-life of $^{150}$Nd and background contamination of the Nd foils have been measured in a previous analysis~\cite{2v2b}, using a total exposure of $0.19$~kg-years. This dataset was used to search for $0\nu4\beta$. 

The signal selection requires three or four tracks with a reconstructed vertex on the $^{150}$Nd source foil (a single composite foil), with a 8~cm maximum vertical spread. Of these tracks, at least three of them have to be associated to a calorimeter hit with an energy of at least 150~keV, and this reconstructed object is an electron candidate. Three independent topological channels are defined: a `four electron' channel (four tracks each with calorimeter hits), a `three electron' channel (three tracks each associated to a calorimeter hit), and a `three electron, one track' channel (four tracks of which only three have calorimeter hits); other topologies have high background contamination. Each electron candidate has to hit a unique calorimeter block. No unassociated calorimeter hits of energy greater than 150~keV (indicative of gamma emission) are allowed. The calorimeter energy and timing is used to exclude electrons that originate in a calorimeter block and cross the source foil.

The three-electron channel has the highest signal efficiency, due to the relatively low energy of the four electrons making it likely that one of the electrons is stopped in the foil. This channel however has the highest background contamination of the three channels, mainly due to SM double beta decays where one of the two beta electrons kicks another atomic electron out of the source foil by M{\o}ller scattering, resulting in three electrons tracks. Other backgrounds due to radioactive contamination are small.
\begin{table}[tbp]
\begin{center}
\begin{tabular}{lcc}  
  Channel & Efficiency (\%) & Background \\ \hline
  Four electron & 0.20 & 16.8  \\
  Three electron & 3.55 & 0.0386 \\
  Three electron, one track & 0.86 & 0.289  
\end{tabular}
  \caption{The signal selection efficiencies (for the best case electron kinematics) and the expected background counts in the $[1.2-2.0]$~MeV window for the three analysis channels.}
\label{tab:effbg}
\end{center}
\end{table}
The overall efficiency and background contamination in the summed-energy window of $[1.2-2.0]$~MeV is shown in Table~\ref{tab:effbg}; the overall efficiency is around $5\%$. 

\begin{figure}[htb]
\centering
\includegraphics[width=0.48\linewidth]{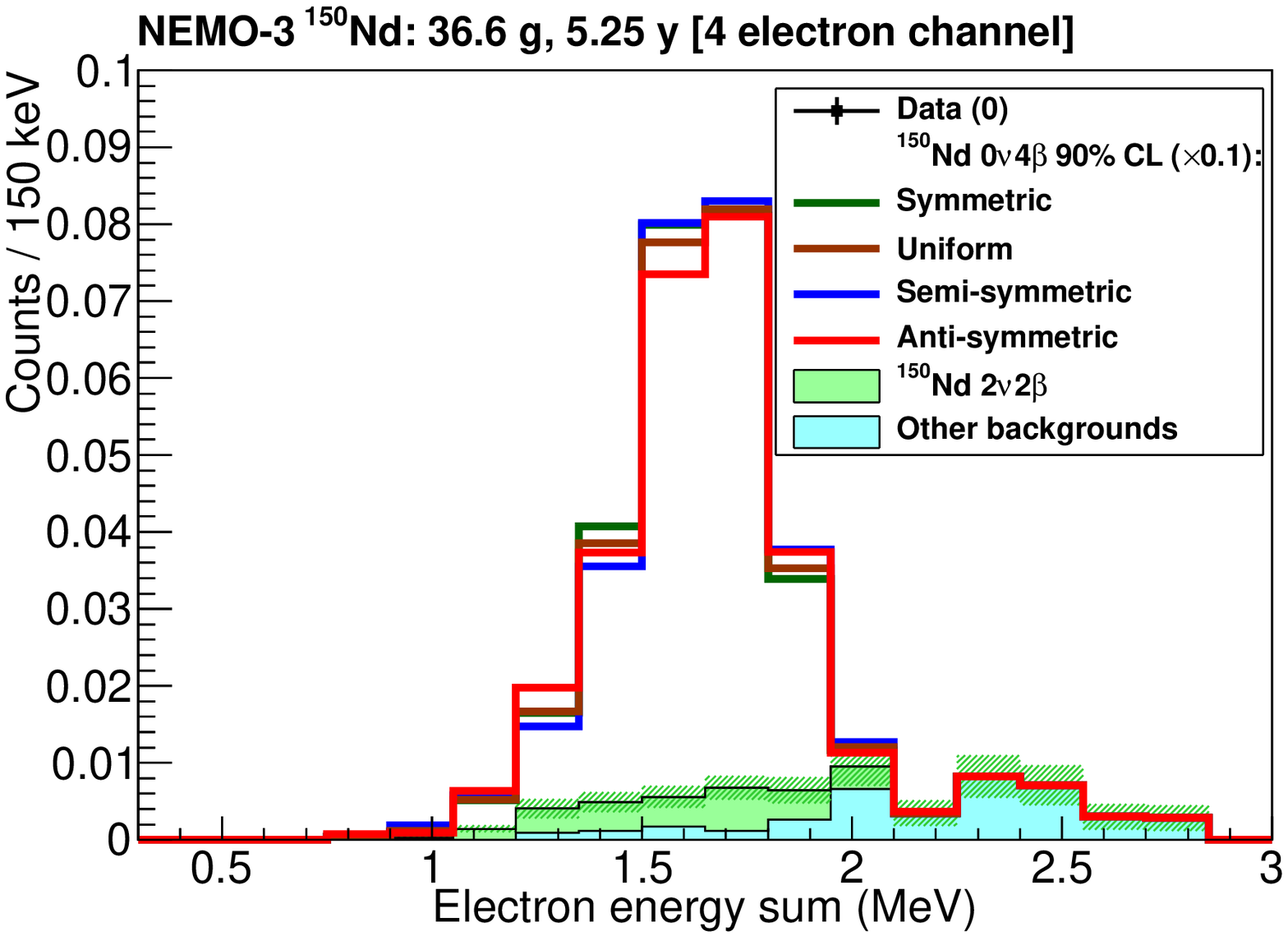}
\includegraphics[width=0.48\linewidth]{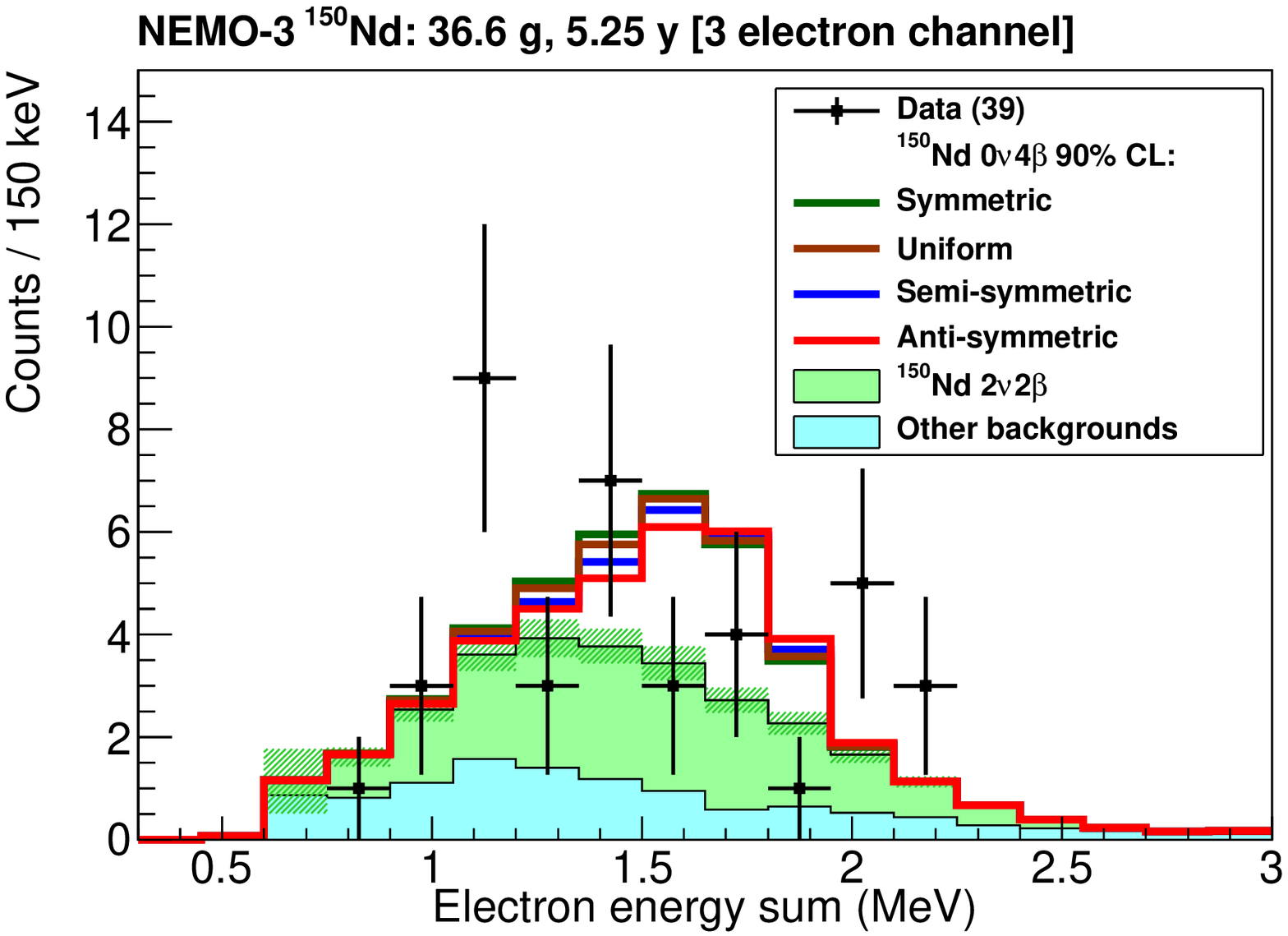}
\includegraphics[width=0.48\linewidth]{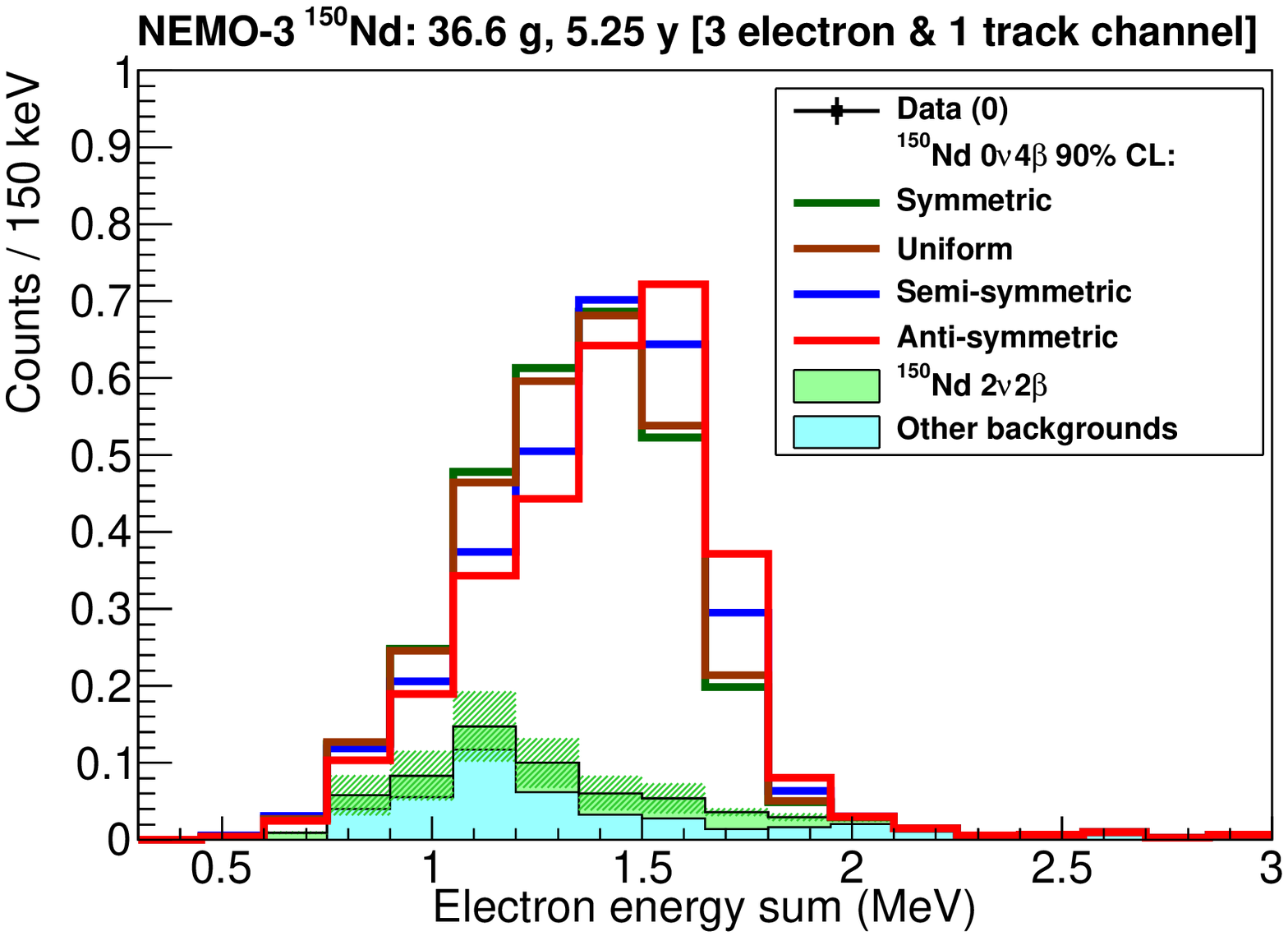}
\hspace{0.0675\textwidth}
{\setlength{\fboxsep}{0pt}\fbox{\includegraphics[height=0.325\linewidth]{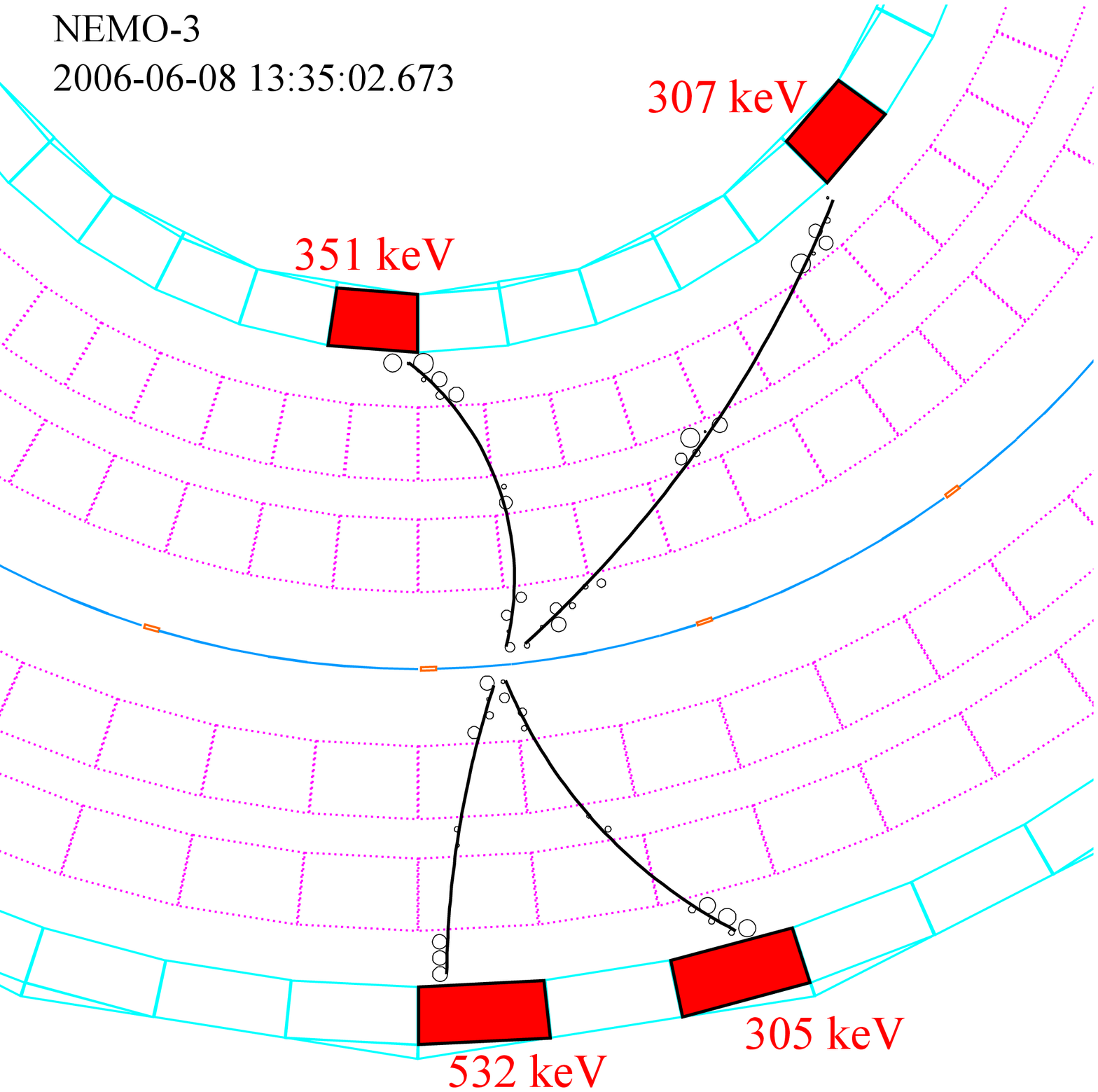}}}
\hspace{0.0675\textwidth}
  \caption{Reconstructed energy spectra for the three different channels, and the expected signal at the 90\% CL limit. Four different electron kinematic distributions are simulated (labelled symmetric, semi-symmetric, anti-symmetric and uniform; described in more detail in Ref.~\cite{prl}). Bottom right: event display of a four electron candidate event in the $^{100}$Mo source foil.}
\label{fig:spectrum}
\end{figure}

As a cross check, the selection is applied to the signal-free $^{100}$Mo source foils, and the event yields in the three channels are consistent with the expected background. In particular, there are two candidate four electron events with an expectation of $2.3\pm0.5$ counts. An event display of one of these four electron candidates is shown in Figure~\ref{fig:spectrum} (bottom right) showing clearly the four reconstructed electrons.

The measured spectra for the three channels are shown in Figure~\ref{fig:spectrum}, and no evidence of this decay is observed. The 90\% CL lower limit on the $^{150}$Nd $0\nu4\beta$ half-life is set between $1.1\times10^{21}$~years and $3.2\times10^{21}$~years, for different kinematic distributions of the four electrons.

\end{document}

%% file: econfmacros.tex
\def\beq{\begin{equation}}
\def\eeq#1{\label{#1}\end{equation}}
\def\eeqn{\end{equation}}

\def\beqa{\begin{eqnarray}}
\def\eeqa#1{\label{#1}\end{eqnarray}}
\def\eeqan{\end{eqnarray}}

\let\bar=\overbar

\def\Dslash{\not{\hbox{\kern-4pt $D$}}}
\def\dslash{\not{\hbox{\kern-2pt $\del$}}}

\def\msb{{\bar{\ssstyle M \kern -1pt S}}}